# Optimally Solving the MCM Problem Using Pseudo-Boolean Satisfiability

Nuno P. Lopes*†    Levent Aksoy*    Vasco Manquinho*†

José Monteiro*†
* INESC-ID
† IST - TU Lisbon

November 2, 2018

## Abstract

In this report, we describe three encodings of the multiple constant multiplication (MCM) problem to pseudo-boolean satisfiability (PBS), and introduce an algorithm to solve the MCM problem optimally. To the best of our knowledge, the proposed encodings and the optimization algorithm are the first formalization of the MCM problem in a PBS manner. This report evaluates the complexity of the problem size and the performance of several PBS solvers over three encodings.

## 1 Introduction

In several computationally intensive operations, such as Finite Impulse Response (FIR) filters as illustrated in Figure 1, Discrete Cosine Transforms (DCT), and Fast Fourier Transforms (FFT), the same input is multiplied by a set of constant coefficients, an operation known as Multiple Constant Multiplications (MCM). The MCM operation is a central operation and performance bottleneck in many applications such as, digital audio and video processing, wireless communication, and computer arithmetic. Hence, hardwired dedicated architectures are the best option for maximum performance and minimum power consumption.

Since the design of a multiplication operation is expensive in terms of area, delay, and power dissipation in hardware and the constants to be multiplied are known beforehand in MCM, the full-flexibility of a multiplier is not necessary in the implementation of the MCM operation. Hence, constant multiplications are generally realized using addition/subtraction and shifting operations in the shift-adds architecture [16]. Since shifts are free in terms of hardware in the parallel realization of MCM, the MCM problem is defined as finding the minimum number of addition/subtraction operations that implement the constant multiplications. The MCM problem has been proved to be NP-hard [4].

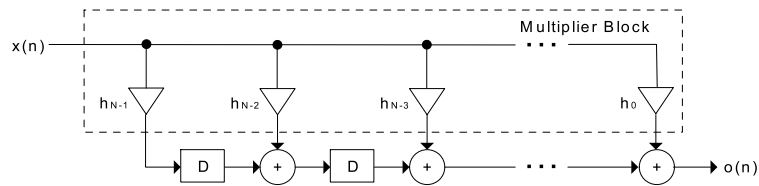

Figure 1: Transposed form of a hardwired FIR filter implementation.

Over the years, many efficient exact and heuristic algorithms based on the pattern search methods [12, 17], 0-1 Integer Linear Programming (ILP) techniques [11, 8, 1], minimum spanning tree algorithms [10], difference-based approaches [15, 9], and graph-based methods [5, 18, 2] have been proposed for the MCM problem. However, in this report, for the first time, we formalize the MCM problem as a PBS problem and find the optimal solution using a generic PBS solver. In this report, we analyze the worst case complexity of the problem and compare the performance of prominent PBS solvers on the MCM problem instances.

The rest of the report is organized as follows: Section 2 gives the background concepts. The encodings of the MCM problem as a PBS problem and the exact algorithm designed for the MCM problem are introduced in Section 3. The experimental results are given in Section 4 and finally, the report is concluded in Section 6.

## 2 Background

This section first gives the problem definition, and then presents an overview on the previously proposed algorithms designed for the MCM problem. At last, the main concepts on PBS and the algorithms designed for the PBS problem are described.

### 2.1 Problem Definition

In the MCM problem, the main operation, called *A-operation* in [18], is an operation with two integer inputs and one integer output that performs a single addition or a subtraction, and an arbitrary number of shifts. It is defined as follows:

$$w = A(u, v) = |(u \ll l_1) + (-1)^s (v \ll l_2)| \gg r = |2^{l_1} u + (-1)^s 2^{l_2} v| 2^{-r} \quad (1)$$

where $l_1, l_2 \geq 0$ are integers denoting left shifts, $r \geq 0$ is an integer indicating the right shift, and $s \in \{0, 1\}$ is the sign that denotes the addition/subtraction operation to be performed.

In the MCM problem, it is assumed that the shifting operation has no cost, since shifts can be implemented only with wires in hardware. Also, the sign of the constant is assumed to be adjusted at some part of the design and the complexity of an adder and a subtracter is equal in hardware. Thus, in the MCM problem, only positive and odd constants are considered. Observe from Eqn. (1) that in the implementation of an odd constant considering any two odd constants at the inputs of an *A-operation*, one of the left shifts, $l_1$ or $l_2$, is zero and $r$ is zero, or both $l_1$ and $l_2$ are zero and $r$ is greater than zero. Thus, the MCM problem can be also defined as follows:

**Definition 1** THE MCM PROBLEM. *Given the target set, $T = \{t_1, \ldots, t_n\} \subset \mathbb{N}$, including the positive and odd unrepeated target constants to be implemented,* find the smallest ready set $R = \{r_0, r_1, \ldots, r_m\}$ with $T \subset R$ such that $r_0 = 1$ and for all $r_k$ with $1 \leq k \leq m$, there exist $r_i, r_j$ with $0 \leq i, j < k$ and an A-operation $r_k = A(r_i, r_j)$.

Hence, the number of operations required to be implemented for the MCM problem is $|R| - 1$ as given in [18]. Thus, to find the minimum number of operations solution of the MCM problem, one has to find the minimum number of intermediate constants such that all the constants, target and intermediate, are implemented using a single *A-operation* where its inputs are '1', intermediate, or target constants, and the MCM implementation is represented in a directed acyclic graph.

## 2.2 Related Work

A straightforward way for the multiplierless realization of constant multiplications, generally known as the digit-based recoding method [7], is to define the constants in multiplications in binary representation and, for each '1' in the binary representation of the constant, shift the variable and add up the shifted variables. As a simple example, consider the constant multiplications $29x = (11101)_{bin}x$ and $43x = (101011)_{bin}x$. The decompositions of constant multiplications are given as follows:

$$29x = (11101)_2 x = x \ll 4 + x \ll 3 + x \ll 2 + x$$
$$43x = (101011)_2 x = x \ll 5 + x \ll 3 + x \ll 1 + x$$

where 6 addition operations are required as illustrated in Figure 2(a). To further improve the solution, one can also define the constants under Canonical Signed Digit (CSD) representation, where each constant has a unique representation with the minimum number of non-zero digits [3].

However, the algorithms that maximize the partial product sharing find the most promising solutions to the MCM problem. They are generally categorized in two classes: the Common Subexpression Elimination (CSE) algorithms [12, 17, 8], and the graph-based (GB) methods [5, 18, 2]. The CSE algorithms, that are also referred to as the pattern search methods, initially define the constants under a particular number representation, *e.g.*, binary, CSD, or Minimal Signed Digit (MSD), and then recursively find the "best" subexpression, generally the most common. The 0-1 ILP formalizations of the MCM problem when the constants are defined under a number representation were first introduced in [11, 8]. However, as shown in [8], the size of the 0-1 ILP problem, *i.e.*, the number of variables, constraints, and optimization variables, grows exponentially with the number of non-zero digits in the representation of a constant. Although problem reduction techniques, that significantly reduce the size of the 0-1 ILP problem and consequently, decrease the CPU time required for a generic 0-1 ILP solver to find a solution, have been proposed in [1], there are MCM instances that exact CSE algorithm cannot handle in a reasonable time as shown in [1]. For our example, suppose that the constants in multiplications are defined in binary. The exact CSE algorithm [1] identifies the most common partial products $3x = (11)_2 x$ and $5x = (101)_2 x$ in both multiplications and obtains a solution with 4 operations as illustrated in Figure 2(b).

On the other hand, the GB algorithms are not limited to any particular number representation and consider a larger number of alternative implementations of a constant multiplication, yielding better solutions than the CSE algorithms, as shown in [18, 2]. The exact GB algorithms that search the fewest number of intermediate constants in breadth-first and depth-first search manners were introduced in [2]. Returning to our example, the exact GB algorithm [2] finds a solution with 3 operations, $7x = x \ll 3 - x$, $29x = 7x \ll 2 + x$, and $43x = 7x \ll 1 + 29x$, as given in Figure 2(c).

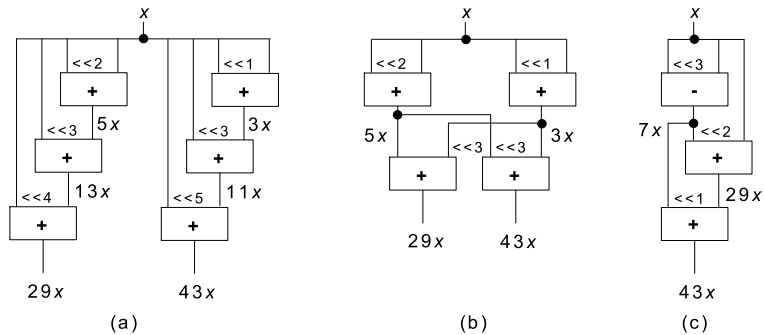

Figure 2: Solutions of the algorithms for the shift-adds implementation of constant multiplications $29x$ and $43x$: (a) the digit-based recoding technique [7]; (b) the CSE method [1]; (c) the graph-based algorithm [2].

## 2.3 Pseudo-Boolean Satisfiability

Recent advances in algorithms for boolean satisfiability (SAT) have led to a significant increase in the capacity and applicability of SAT solvers. One of these applications is the pseudo-boolean satisfiability (PBS) and optimization (PBO), also called 0-1 ILP (integer linear programming).

PBS constraints are of the form:

$$\sum_{i=1}^{N} c_i x_i \geq b$$

where each variable $x_i$ may only take the values 0 or 1, and $b$ and each $c_i$ are arbitrary integers (including negative integers). In PBO, there is an additional objective clause whose value should be minimized.

# 3  The Algorithm

In this section, we describe the formalization of the MCM problem as a PBS problem. Unlike the exact CSE algorithms [11, 8], that extract all the possible implementations of a constant under a particular number representation, our algorithm is not limited to any particular number representation and can consider the search space of a GB algorithm.

We first describe the encoding of several auxiliary operations that are used in the encodings of the MCM problem as a PBS problem. Then, we introduce three different encodings, give their worst case analysis, and present the optimization algorithm. At last, further possible reductions on the size of the problem are described.

We use the big-endian encoding for the numbers. For example, 10 is represented as (assuming 4-bits numbers): $x_1 = 1 \land x_2 = 0 \land x_3 = 1 \land x_4 = 0$.

## 3.1 Auxiliary procedures

### 3.1.1 2-operand XOR

The first operation is an XOR of two operands, with the result assigned to the first variable. Formally, let the function $\mathsf{XOR}(a, b, c)$ be:

$a \Leftrightarrow (b\ \mathbf{XOR}\ c)$

$$\begin{aligned}
-a + b + c &\geq 0 \quad \land \\
-a - b - c &\geq -2 \quad \land \\
a + b - c &\geq 0 \quad \land \\
a - b + c &\geq 0
\end{aligned}$$

### 3.1.2 Conditional 2-operand XOR

The second operation is also an XOR of two operands, but the result is only assigned to the result variable if the last variable (the condition) is true. Formally, let the function $\mathsf{cond\_XOR}(a, b, c, d)$ be:

$d \Rightarrow (a \Leftrightarrow (b\ \mathbf{XOR}\ c))$

$$\begin{aligned}
-d - a + b + c &\geq -1 \quad \land \\
-d - a - b - c &\geq -3 \quad \land \\
-d + a + b - c &\geq -1 \quad \land \\
-d + a - b + c &\geq -1
\end{aligned}$$

### 3.1.3 3-operand XOR

This function is similar with the first one, but it computes the XOR of three variables. Formally, let the function $\mathsf{XOR}(a, b, c, d)$ be:

$a \Leftrightarrow (b\ \mathbf{XOR}\ c\ \mathbf{XOR}\ d)$

$$\begin{aligned}
-a + b + c + d &\geq 0 \quad \land \\
-a + b - c - d &\geq -2 \quad \land \\
-a - b + c - d &\geq -2 \quad \land \\
-a - b - c + d &\geq -2 \quad \land \\
a - b - c - d &\geq -2 \quad \land \\
a - b + c + d &\geq 0 \quad \land \\
a + b - c + d &\geq 0 \quad \land \\
a + b + c - d &\geq 0
\end{aligned}$$

### 3.1.4 Conditional 3-operand XOR

This function is similar with the second one, but it conditionally computes the XOR of three variables. Formally, let the function $\mathsf{cond\_XOR}(a, b, c, d, e)$ be:

$e \Rightarrow (a \Leftrightarrow (b \text{ XOR } c \text{ XOR } d))$

$$\begin{aligned}
-e - a + b + c + d &\geq -1 \quad \wedge \\
-e - a + b - c - d &\geq -3 \quad \wedge \\
-e - a - b + c - d &\geq -3 \quad \wedge \\
-e - a - b - c + d &\geq -3 \quad \wedge \\
-e + a - b - c - d &\geq -3 \quad \wedge \\
-e + a - b + c + d &\geq -1 \quad \wedge \\
-e + a + b - c + d &\geq -1 \quad \wedge \\
-e + a + b + c - d &\geq -1
\end{aligned}$$

### 3.1.5 Conditional Bit Copy

This function copies one bit if the condition flag (first parameter) is true. Formally, let $\mathsf{cond\_copy}(a, b, c)$ be:

$a \Rightarrow (b \Leftrightarrow c)$

$$\begin{aligned}
-a - b + c &\geq -1 \quad \wedge \\
-a + b - c &\geq -1
\end{aligned}$$

### 3.1.6 N-Bits Adder

This function implements a full adder with ripple carry, which stores the result of the addition in the first argument. The variables $d_{1..N-1}$ are fresh (i.e., they are created in this function and they are not used elsewhere). $d_i$ is the carry bit used to compute the $i$-th result bit and therefore it is computed with the $(i+1)$-th operands and carry bits. Formally, let $a \leftarrow \mathsf{adder}(b, c)$ be:

$a_{1..N} \Leftrightarrow b_{1..N} + c_{1..N}$

$$\begin{aligned}
&(* \text{ compute the carry bits } *) \\
&\bigwedge_{i=1..N-2} -2d_i + b_{i+1} + c_{i+1} + d_{i+1} \geq 0 \quad \wedge \\
&\bigwedge_{i=1..N-2} 2d_i - b_{i+1} - c_{i+1} - d_{i+1} \geq -1 \quad \wedge \\
&\qquad -2d_{N-1} + b_N + c_N \geq 0 \quad \wedge \\
&\qquad d_{N-1} - b_N - c_N \geq -1 \quad \wedge \\
&(* \text{ disallow overflows } *) \\
&\qquad -b_1 - c_1 - d_1 \geq -1 \quad \wedge \\
&(* \text{ and do the addition } *) \\
&\bigwedge_{i=1..N-1} \mathsf{XOR}(a_i, b_i, c_i, d_i) \quad \wedge \\
&\qquad \mathsf{XOR}(a_N, b_N, c_N)
\end{aligned}$$

### 3.1.7 Conditional N-Bits Adder

This function also implements a full adder, but the result of the operation is only assigned if the last argument (the condition) is true. The variables $d_{1..N-1}$ (the carry bits) are fresh. Formally, let $a \leftarrow \mathsf{cond\_adder}(b, c, e)$ be:

$$e \Rightarrow (a_{1..N} \Leftrightarrow b_{1..N} + c_{1..N})$$

$$
\begin{aligned}
& (* \text{ compute the carry bits } *) \\
& \bigwedge_{i=1..N-2} -2d_i + b_{i+1} + c_{i+1} + d_{i+1} \geq 0 \quad \wedge \\
& \bigwedge_{i=1..N-2} 2d_i - b_{i+1} - c_{i+1} - d_{i+1} \geq -1 \quad \wedge \\
& \qquad -2d_{N-1} + b_N + c_N \geq 0 \quad \wedge \\
& \qquad d_{N-1} - b_N - c_N \geq -1 \quad \wedge \\
& (* \text{ disallow overflows } *) \\
& \qquad -e - b_1 - c_1 - d_1 \geq -2 \quad \wedge \\
& (* \text{ and do the addition } *) \\
& \bigwedge_{i=1..N-1} \text{cond\_XOR}(a_i, b_i, c_i, d_i, e) \quad \wedge \\
& \qquad \text{cond\_XOR}(a_N, b_N, c_N, e)
\end{aligned}
$$

### 3.1.8 Conditional N-Bits Adder with Conditional Carry

This function also implements a conditional full adder, with the difference being that the carry bits may not be fresh (e.g., they can be shared with other conditional adders/subtractors, provided that at most one of them is enabled). Formally, let $a \leftarrow \text{cond\_adder}(b, c, d, e)$ be:

$$e \Rightarrow (a_{1..N} \Leftrightarrow b_{1..N} + c_{1..N})$$

$$
\begin{aligned}
& (* \text{ compute the carry bits } *) \\
& \bigwedge_{i=1..N-2} -2e - 2d_i + b_{i+1} + c_{i+1} + d_{i+1} \geq -2 \quad \wedge \\
& \bigwedge_{i=1..N-2} -2e + 2d_i - b_{i+1} - c_{i+1} - d_{i+1} \geq -3 \quad \wedge \\
& \qquad -2e - 2d_{N-1} + b_N + c_N \geq -2 \quad \wedge \\
& \qquad -e + d_{N-1} - b_N - c_N \geq -2 \quad \wedge \\
& (* \text{ disallow overflows } *) \\
& \qquad -e - b_1 - c_1 - d_1 \geq -2 \quad \wedge \\
& (* \text{ and do the addition } *) \\
& \bigwedge_{i=1..N-1} \text{cond\_XOR}(a_i, b_i, c_i, d_i, e) \quad \wedge \\
& \qquad \text{cond\_XOR}(a_N, b_N, c_N, e)
\end{aligned}
$$

### 3.1.9 N-Bits Subtractor

This function implements a subtractor, similar with the first adder. The variables $d_{1..N-1}$ are the borrow bits and they are fresh. Formally, let $a \leftarrow \text{subtractor}(b, c)$ be:

$$a_{1..N} \Leftrightarrow b_{1..N} - c_{1..N}$$

$$(* \text{ compute the borrow bits } *)$$

$$\bigwedge_{i=1..N-2} 2d_i + b_{i+1} - c_{i+1} - d_{i+1} \geq 0 \quad \wedge$$

$$\bigwedge_{i=1..N-2} -2d_i - b_{i+1} + c_{i+1} + d_{i+1} \geq -1 \quad \wedge$$

$$-2d_{N-1} - b_N + c_N \geq -1 \quad \wedge$$

$$d_{N-1} + b_N - c_N \geq 0 \quad \wedge$$

(∗ disallow overflows ∗)

$$b_1 - c_1 \geq 0 \quad \wedge$$
$$b_1 - d_1 \geq 0 \quad \wedge$$
$$-c_1 - d_1 \geq -1 \quad \wedge$$

(∗ and do the subtraction ∗)

$$\bigwedge_{i=1..N-1} \mathsf{XOR}(a_i, b_i, c_i, d_i) \quad \wedge$$

$$\mathsf{XOR}(a_N, b_N, c_N)$$

### 3.1.10 Conditional N-Bits Subtractor

This function implements a conditional subtractor, similar with the second adder. The variables $d_{1..N-1}$ are the borrow bits and they are fresh. Formally, let $a \leftarrow \mathsf{cond\_subtractor}(b, c, e)$ be:

$$e \Rightarrow (a_{1..N} \Leftrightarrow b_{1..N} - c_{1..N})$$

(∗ compute the borrow bits ∗)

$$\bigwedge_{i=1..N-2} 2d_i + b_{i+1} - c_{i+1} - d_{i+1} \geq 0 \quad \wedge$$

$$\bigwedge_{i=1..N-2} -2d_i - b_{i+1} + c_{i+1} + d_{i+1} \geq -1 \quad \wedge$$

$$-2d_{N-1} - b_N + c_N \geq -1 \quad \wedge$$

$$d_{N-1} + b_N - c_N \geq 0 \quad \wedge$$

(∗ disallow overflows ∗)

$$-e + b_1 - c_1 \geq -1 \quad \wedge$$
$$-e + b_1 - d_1 \geq -1 \quad \wedge$$
$$-e - c_1 - d_1 \geq -2 \quad \wedge$$

(∗ and do the subtraction ∗)

$$\bigwedge_{i=1..N-1} \mathsf{cond\_XOR}(a_i, b_i, c_i, d_i, e) \quad \wedge$$

$$\mathsf{cond\_XOR}(a_N, b_N, c_N, e)$$

### 3.1.11 Conditional N-Bits Subtractor with Conditional Borrow

This function implements a conditional subtractor, similar with the third adder. The variables $d_{1..N-1}$ are the borrow bits and they may not be fresh. Formally, let $a \leftarrow \mathsf{cond\_subtractor}(b, c, d, e)$ be:

$$e \Rightarrow (a_{1..N} \Leftrightarrow b_{1..N} - c_{1..N})$$

(∗ compute the borrow bits ∗)

$$\bigwedge_{i=1..N-2} -2e + 2d_i + b_{i+1} - c_{i+1} - d_{i+1} \geq -2 \quad \wedge$$

$$\bigwedge_{i=1..N-2} -2e - 2d_i - b_{i+1} + c_{i+1} + d_{i+1} \geq -3 \quad \wedge$$

$$-2e - 2d_{N-1} - b_N + c_N \geq -3 \quad \wedge$$

$$-e + d_{N-1} + b_N - c_N \geq -1 \quad \wedge$$

(∗ disallow overflows ∗)

$$-e + b_1 - c_1 \geq -1 \quad \wedge$$

$$-e + b_1 - d_1 \geq -1 \quad \wedge$$

$$-e - c_1 - d_1 \geq -2 \quad \wedge$$

(∗ and do the subtraction ∗)

$$\bigwedge_{i=1..N-1} \mathsf{cond\_XOR}(a_i, b_i, c_i, d_i, e) \quad \wedge$$

$$\mathsf{cond\_XOR}(a_N, b_N, c_N, e)$$

### 3.1.12 Conditional Left Shift

This function computes the left shift of the third parameter by a given constant. The result is only assigned if the condition flag (first parameter) is true. Formally, let $\mathsf{cond\_shift\_left}(a, b, c, C)$ be:

$a \Rightarrow (b_{1..N} \Leftrightarrow (c_{1..N} \ll C))$, with $C \in [0, N-1]$.

(∗ zero out last C bits ∗)

$$\bigwedge_{i=(N-C+1)..N} -a - b_i \geq -1 \quad \wedge$$

(∗ copy remaining bits ∗)

$$\bigwedge_{i=1..(N-C)} \mathsf{cond\_copy}(a, b_i, c_{i+C}) \quad \wedge$$

(∗ disallow overflows ∗)

$$\bigwedge_{i=1..C} -a - c_i \geq -1 \quad \wedge$$

### 3.1.13 $N$ Bits Left Shifts

This function computes the left shift of the second parameter by some constant, i.e., this function can return the second parameter shifted left by any amount in the range $[0, N-1]$. The variables $c_{1..N}$ are fresh, and they are used to restrict the solution so that only one shift unit is enabled. Formally, let $a \leftarrow \mathsf{shift\_left}(b)$ be:

$\exists C \in [0, N-1] : (a_{1..N} \Leftrightarrow b_{1..N} \ll C)$

$$\sum_{i=1}^{N} c_i = 1 \quad \wedge$$

$$\bigwedge_{i=0..(N-1)} \mathsf{cond\_shift\_left}(c_{i+1}, a, b, i)$$

### 3.1.14 Conditional Right Shift

This function computes the right shift of the third parameter by a given constant. The result is only assigned if the condition flag (first parameter) is true. Formally, let $\mathsf{cond\_shift\_right}(a, b, c, C)$ be:
$a \Rightarrow (b_{1..N} \Leftrightarrow (c_{1..N} \gg C))$, with $C \in [0, N-1]$.

$$\begin{array}{cc}
& (* \text{ zero out first C bits } *) \\
\bigwedge_{i=1..C} & -a - b_i \geq -1 \quad \wedge \\
& (* \text{ copy remaining bits } *) \\
\bigwedge_{i=1..(N-C)} & \mathsf{cond\_copy}(a, b_{i+C}, c_i) \quad \wedge \\
& (* \text{ disallow overflows } *) \\
\bigwedge_{i=(N-C+1)..N} & -a - c_i \geq -1 \quad \wedge
\end{array}$$

### 3.1.15 $N$ Bits Right Shifts

The $a \leftarrow \mathsf{shift\_right}(b)$ function is similar with $\mathsf{shift\_left}$:
$\exists C \in [0, N-1] : (a_{1..N} \Leftrightarrow b_{1..N} \gg C)$

$$\begin{array}{cc}
& \sum_{i=1}^{N} c_i = 1 \quad \wedge \\
\bigwedge_{i=0..(N-1)} & \mathsf{cond\_shift\_right}(c_{i+1}, a, b, i)
\end{array}$$

### 3.1.16 Equate Variable List to Variable

This function takes a list of variables and returns one of them. The variables $x_{1..|V|}$ are fresh and are used to restrict the solution so that only one list member is returned. Formally, let $a \leftarrow \mathsf{equate\_var\_list\_to\_var}(V)$ be:
$\exists v \in V : (a_{1..N} \Leftrightarrow v_{1..N})$

$$\sum_{i=1}^{|V|} x_i >= 1 \quad \wedge$$
$$\forall v_j \in V : \quad \bigwedge_{i=1..N} \mathsf{cond\_copy}(x_j, v_{ji}, a_i)$$

### 3.1.17 Equate Variable List to Constant

This function is similar to the previous one, but it matches the variable list against a constant, and therefore an optimized encoding can be produced. The variables $x_{1..|V|}$ are fresh. Formally, let $\mathsf{equate\_var\_list\_to\_const}(V, C)$ be:
$\exists v \in V : (C_{1..N} \Leftrightarrow v_{1..N})$

$$\sum_{i=1}^{|V|} x_i >= 1 \quad \wedge$$
$$\forall \quad v_j \in V : \bigwedge_{i=1..N} C_i = 1 \Rightarrow -x_j + v_{ji} \geq 0 \quad \wedge$$
$$\forall \quad v_j \in V : \bigwedge_{i=1..N} C_i = 0 \Rightarrow -x_j - v_{ji} \geq -1$$

Since C is a constant, the implications in the above formula are folded away when encoding.

## 3.2 Encoding Algorithm

In this section we describe an algorithm that generates a PBS problem that is equivalent to the problem of finding a solution of an MCM problem with a fixed number of addition/subtraction operations. The algorithm is shown in Fig. 3.

The algorithm takes as input the set of positive, odd, and unrepeated target constants, and the exact number of addition/subtraction operations required to implement the multiplications (A-operations). The idea is to assign the output of each possible operation to an intermediate variable. Using this set of variables, we add additional constraints so that a final variable takes the value of one of the other variables. In the end, we have $x$ of these final variables (where $x$ is the number of operations). We then just need to constrain the solution so that at least one of these variables equates each target constant.

The number of bits of each variable, $N$, is defined as the number of bits of the largest target constant plus one.

The algorithm works as follows. In lines 2–13, we generate all the possible candidate operations for each intermediate operation. The possible candidate operations are all the combinations of additions and subtractions of the input variable (possibly left shifted), and the previously generated intermediate operations. In line 14, we generate a constraint to assign the value of one of the candidates (possibly right shifted) to the intermediate variable.

Finally, in lines 15–19, we constrain the solution so that at least one intermediate variable (possibly left shifted) is equal to each constant.

All the auxiliary functions were defined in the previous section, except the exactly($x$) function, which creates (and returns) a variable of $N$ bits so that exactly $x$ bits are true. For example, exactly(1) represents the set of the powers of two up to $2^{N-1}$. This function has a trivial encoding in PBS.

### 3.2.1 Alternative Encodings

The encoding given in Figure 3 is our first encoding. We have, however, implemented alternative encodings. The other two encodings are generated with the same algorithm, except for the following differences:

- Encoding 2 uses conditional adders/subtractors

- Encoding 3 uses conditional adders/subtractors with conditional carry/borrow

- Line 14 of the algorithm is deleted in both encodings

- exactly(1) and left shifts are computed just once per intermediate variable in both encodings (i.e., their results are reused by the several adders/subtractors), except in line 5 where we need to compute a second exactly(1)

The reuse of the results of the shifts and the exactly constraints are possible since the addition and subtraction operations are made conditional. They all output the result to the same variable (hence line 14 becomes useless). However, they are restricted so that only one of them can produce the output.

```
      procedure EncodeMCM
      input
          consts ⊂ ℕ
          operations ∈ ℕ
      vars
          f - final PBS formula
          v - list of alternative operations
          M - map to store the variables of each operation
      begin
 1        f := true
 2        for i = 1..operations do
 3            v := ∅
 4            v := v ∪ {exactly(2)}
 5            v := v ∪ {subtractor(exactly(1), exactly(1))}
 6            for op1 = 1..(i − 1) do
 7                v := v ∪ {adder(left_shift(M[op1]), exactly(1))}
 8                v := v ∪ {subtractor(left_shift(M[op1]), exactly(1))}
 9                v := v ∪ {subtractor(exactly(1), left_shift(M[op1]))}
10                for op2 = op1..(i − 1) do
11                    v := v ∪ {adder(left_shift(M[op1]), left_shift(M[op2]))}
12                    v := v ∪ {subtractor(left_shift(M[op1]), left_shift(M[op2]))}
13                    v := v ∪ {subtractor(left_shift(M[op2]), left_shift(M[op1]))}
14            M[i] := right_shift(equate_var_list_to_var(v))

15        for each c ∈ consts do
16            v := ∅
17            for i = 1..operations do
18                v := v ∪ left_shift(M[i])
19            f := f ∧ equate_var_list_to_const(v, c)
20        return f
      end
```

Figure 3: MCM encoding algorithm.

These alternative encodings reduce the number of variables at the expense of bigger constraints. They also reduce considerably the number of constraints, due to the sharing of the results of the shift operations.

#### 3.2.2 Complexity of the Encoding

The third encoding (the smallest in terms of the number of variables and constraints) has the following complexity:

- Number of variables: $O(A^3 N)$
- Number of constraints: $O(A^3 N^2)$

where $A$ is the number of A-operations, and $N$ is the number of bits. However, in the worst case, the size of the search space is still exponential in terms of $A$ and $N$.

These bounds can be obtained by observing the following:

1. Each iteration of the outer loop generates $O(i^2)$ addition/subtraction operations, as well as left shifts (as the loop goes through all the possible combinations of additions and subtractions of the previous intermediate constants).

2. The outer loop executes $A$ times, and thus we get a bound of $O(A^3)$ in the number of addition, subtraction, and left shift operations.

3. Each addition and subtraction operation requires $O(N)$ variables and constraints.

4. Each left shift operation requires $O(N)$ variables and $O(N^2)$ constraints.

5. The number of target constants is at most $A$, and thus the number of operations generated by the loop of lines 15–19 is bounded by $O(A^2)$.

### 3.3 Optimization Algorithm

In this section we describe an optimization algorithm that yields the optimal solution for a given MCM problem. The algorithm is shown in Fig. 4.

```
    procedure OptimalMCM
    input
        consts ⊂ ℕ
        upperbound ∈ ℕ
    begin
1       upperbound := upperbound − 1
2       while EncodeMCM(consts, upperbound) is SAT do
3           upperbound := upperbound − 1
4       return upperbound + 1
    end
```

Figure 4: MCM optimization algorithm.

The OptimalMCM procedure takes two arguments: the set of target constants, and an upper bound on the number of required addition/subtraction operations. This bound can be efficiently computed by an approximate algorithm, such as the ones described in Section 2.2. The only requirement on the upper bound is that there must exist a solution to the problem with the given number of operations.

The algorithm starts by trying to find the solution with one fewer operation than the upper bound (line 1), since we already know that there exists a solution with the upper bound number of operations. Then it decreases the maximum number of operations until no solution exists (and thus the PBS solver returns UNSAT). When this occurs, the algorithm returns the last solution found by the solver.

### 3.4 Further Improvements

We now describe several optimizations that can be implemented to improve the PBS encodings given before, and thus the overall running time:

1. Constrain the result of the subtractor of line 5 so that the result is non-zero. This can be encoded as $-a - b \geq -1$ for every pair of input bits.

2. Left shift of line 18 can be omitted if the constant is odd, since the result of a non-zero left shift is always an even number.

3. Compute all possible constants generated by one operation before encoding (computationally inexpensive). If any target constant can be implemented by a single operation, then the number of possible operations is reduced by one (since one operation is generating that constant). Unit propagation can then be applied in the shifts that take the constant as input. This procedure can be executed iteratively by combining all the target constants that can be implemented with a single operation.

4. If the optimization step 3 is implemented, then lines 4 and 5 can be executed only in the first $x$ iterations of the loop, where $x$ is the number of intermediate constants (maximum number of operations minus the number of target constants). This optimization is possible since the previous optimization will remove all the target constants that can be implemented with a single operation.

5. If there are no target constants implementable with a single operation, then the value of the variable $i$ in the loop of line 17 can start from 2.

## 4 Evaluation

### 4.1 Random Tests

In this section we evaluate the three proposed encodings with different PBS solvers (bsolo [13], minisat+ [6], and wbo [14]). We generated a random set of tests with different number of bits and target constants. We also added a test for a FIR filter (tests 10/11). The number of A-operations, the target constants, and the satisfiability (if known) of each test is shown in Table 1. The PB complexity of each test and encoding in terms of number of clauses and number of variables is show in Table 2. Our implementation does not consider right shifts, since they are usually not necessary in order to achieve an optimal solution, and because they increase the search space.

| Test | # Ops | Target Constants | SAT |
|---|---|---|---|
| 01 | 5 | 731951 | ✓ |
| 02 | 4 | 731951 | × |
| 03 | 5 | 33951 | ✓ |
| 04 | 3 | 33951 | × |
| 05 | 12 | 15783; 47351; 1345; 111111; 9871 | ✓ |
| 06 | 11 | 15783; 47351; 1345; 111111; 9871 | ? |
| 07 | 15 | 1571; 3579; 7777; 1351; 123; 9999; 135; 767 | ✓ |
| 08 | 13 | 1571; 3579; 7777; 1351; 123; 9999; 135; 767 | ✓ |
| 09 | 12 | 1571; 3579; 7777; 1351; 123; 9999; 135; 767 | ? |
| 10 | 17 | 1701; 709; 1015; 1269; 1203; 683; 201; 565; 1653; 681; 17; 261; 4621; 3435 | ✓ |
| 11 | 16 | 1701; 709; 1015; 1269; 1203; 683; 201; 565; 1653; 681; 17; 261; 4621; 3435 | ? |

Table 1: Details (number of A-operations, target constants, and known satisfiability) of the benchmarks.

| Test | Encoding 1 | | | Encoding 2 | | | Encoding 3 | | |
|---|---|---|---|---|---|---|---|---|---|
| | # Clauses | # Vars | File | # Clauses | # Vars | File | # Clauses | # Vars | File |
| 01 | 156,096 | 11,243 | 4.7 | 46,243 | 3,620 | 1.7 | 46,243 | 1,840 | 1.7 |
| 02 | 81,762 | 6,075 | 2.5 | 24,920 | 2,043 | 0.89 | 24,920 | 1,103 | 0.87 |
| 03 | 105,936 | 9,103 | 3.2 | 33,299 | 2,932 | 1.3 | 33,299 | 1,508 | 1.2 |
| 04 | 23,837 | 2,214 | 0.71 | 8,027 | 803 | 0.28 | 8,027 | 483 | 0.29 |
| 05 | 1,469,379 | 111,392 | 48 | 417,306 | 32,552 | 17 | 417,306 | 14,685 | 17 |
| 06 | 1,141,533 | 86,924 | 37 | 326,275 | 25,526 | 13 | 326,275 | 11,586 | 13 |
| 07 | 2,012,698 | 175,694 | 68 | 594,926 | 50,759 | 25 | 594,926 | 22,941 | 25 |
| 08 | 1,327,081 | 116,501 | 44 | 395,824 | 33,866 | 16 | 395,824 | 15,414 | 16 |
| 09 | 1,051,800 | 92,642 | 35 | 315,408 | 27,032 | 13 | 315,408 | 12,360 | 13 |
| 10 | 2,557,759 | 234,851 | 88 | 767,018 | 67,439 | 32 | 767,018 | 30,480 | 32 |
| 11 | 2,142,844 | 197,121 | 73 | 644,749 | 56,715 | 27 | 644,749 | 25,684 | 27 |

Table 2: PB complexity of the tests of Table 1 in terms of number of clauses, number of variables, and OPB file size (in MB).

The objectives of this evaluation were: check if the proposed method scales to a realistic number of bits and target constants, measure the time to prove a solution satisfiable and unsatisfiable (so that we could derive an optimization algorithm), and quantify how worse are the approximated solutions when compared with the optimal one.

The running time of each of the solvers over each of the three proposed encodings is shown in Table 3.

First we note that proving UNSAT is considerably slower than proving SAT, and that is why the optimization algorithm we propose searches for the optimal solution starting at the upper bound. Second, we note that finding a solution with a bigger number of operations is faster, since the number of solutions is higher, and thus it increases the chances that the solver will find a solution more quickly.

| Test | Encoding 1 | | | Encoding 2 | | | Encoding 3 | | |
|---|---|---|---|---|---|---|---|---|---|
| | bsolo | minisat+ | wbo | bsolo | minisat+ | wbo | bsolo | minisat+ | wbo |
| 01 | 17727.83 | 104.78 | 285.92 | 35.69 | 9.93 | 5.37 | 796.49 | 3.12 | 4.85 |
| 02 | T/O | 1989.27 | 5003.81 | T/O | 390.56 | 1345.10 | T/O | 292.64 | 463.01 |
| 03 | 32.32 | 10.84 | 6.74 | 0.64 | 1.31 | 0.84 | 2.45 | 1.13 | 0.45 |
| 04 | 69.28 | 2.44 | 3.11 | 14.79 | 1.09 | 2.02 | 16.62 | 1.36 | 1.74 |
| 05 | T/O | T/O | T/O | T/O | T/O | T/O | T/O | T/O | T/O |
| 06 | T/O | T/O | T/O | T/O | T/O | T/O | T/O | T/O | T/O |
| 07 | T/O | T/O | T/O | T/O | T/O | T/O | T/O | T/O | T/O |
| 08 | T/O | T/O | T/O | T/O | T/O | T/O | T/O | T/O | T/O |
| 09 | T/O | T/O | T/O | T/O | T/O | T/O | T/O | T/O | T/O |
| 10 | T/O | T/O | T/O | T/O | T/O | T/O | T/O | T/O | T/O |
| 11 | T/O | T/O | T/O | T/O | T/O | T/O | T/O | T/O | T/O |

Table 3: Time (in seconds) that each PBS solver takes to solve the benchmarks. T/O means that the solver timed out (the limit was twelve hours).

Regarding the different encodings, the third one yields the best results. In terms of solvers, minisat+ solves as many problems as wbo (with the third encoding), but minisat+ scales better.

Finally, we note that no solver scales to the bigger tests (either with a high number of bits or with a high number of intermediate constants).

### 4.2 Random FIR filers

We performed a second set of tests. We automatically generated these tests from 600 randomly generated FIR filters of 10 and 11 bits. We generated two tests per filter, changing only the number of operations: one with the upper bound value (a SAT test), and another with one fewer operations (most likely UNSAT, since approximate algorithms are usually close to the optimal for small numbers).

We run the tests for four weeks, with the three solvers in parallel, and we gave five days of time limit for each test. In this experiment, we only considered the third encoding, since it showed to be the most beneficial and the more succinct (even so, some tests had several thousand of variables and millions of constraints).

The maximum number of tests considered by a solver was 440, but only 124 (28 %) of these were non-trivial. 32 % of the tests were marked as trivially SAT, and 40 % were marked as trivially UNSAT. Trivial tests are marked as such by the encoding program when it can solve the problem itself without needing a solver. Thereby the tests marked as trivial were not run by the solvers, and we do not take them into account in the results.

The non-trivial tests had an average number of clauses of 284,986 (with 67,146 of minimum, and 1,282,690 of maximum), and an average number of variables of 12,808 (with 3,164 of minimum, and 55,296 of maximum).

| Solver | SAT | | UNSAT | | Combined | | |
|---|---|---|---|---|---|---|---|
| | # Tests | Avg. Time | # Tests | Avg. Time | # Tests | Avg. Time | % |
| VBS | 53 | 9,712 s | 28 | 19,280 s | 81 | 13,019 s | 65.3 % |
| bsolo | 19 | 28,835 s | 11 | 28,656 s | 30 | 28,770 s | 24.2 % |
| minisat+ | 30 | 3,017 s | 20 | 25,922 s | 50 | 12,179 s | 40.3 % |
| wbo | 51 | 9,755 s | 26 | 1,730 s | 77 | 7,045 s | 62.1 % |

Table 4: Number of instances solved, and the average solving time that each solver took to give the solution in our FIR filters benchmark.

| Solver | SAT | | UNSAT | | Combined | |
|---|---|---|---|---|---|---|
| | # Tests | % | # Tests | % | # Tests | % |
| VBS | 53 | 42.7 % | 28 | 22.6 % | 81 | 65.3 % |
| bsolo | 0 | 0 % | 0 | 0 % | 0 | 0 % |
| minisat+ | 13 | 10.5 % | 8 | 6.5 % | 21 | 16.9 % |
| wbo | 40 | 32.3 % | 20 | 16.1 % | 60 | 48.5 % |

Table 5: Number and percentage of instances where each solver yielded the best running time in our FIR filters benchmark.

Table 4 shows the number of tests solved by each solver. We also give the average time taken by each solver, plus the overall percentage of solved instances. VBS (virtual best solver) is also shown in the results, which is a theoretical solver that gives the best result obtained by the combination of the three solvers tested. This can be achieved in practice by, for example, running all the solvers in parallel and stop when one of the solvers finishes.

Table 5 shows the number and the percentage of instances where each solver resulted in the best running time. Although there is no absolute winner, wbo is the best overall performer. bsolo, on the other hand, never outperformed the other solvers.

## 5 Acknowledgments

This work was partially supported by FCT under research project iExplain (PTDC/EIA-CCO/102077/2008), the grant SFRH/BD/63609/2009, and INESC-ID multiannual funding through the PIDDAC program funds.

## 6 Conclusions

In this report, we presented, to the best of our knowledge, the first encoding of the MCM problem to PBS. We describe an exact algorithm for the problem, and we evaluate three different solvers on a set of benchmark problems.

Our approach scales only up to about 10/11 bits of the target constants with a moderate number of A-operations. Further optimizations in the encoding and smarter PBS algorithms are still required in order to make our approach scale to more realistic problems.